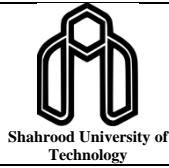

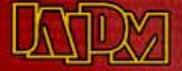

Review Paper

# Game Theory Solutions in Sensor-Based Human Activity Recognition: A Review

Mohammad Hossein Shayesteh, Behrooz Shahrokhzadeh* and Behrooz Masoumi

*Department of Computer and Information Technology Engineering, Qazvin Branch, Islamic Azad University, Qazvin, Iran.*

**Abstract**

The Human Activity Recognition (HAR) tasks automatically identify human activities using the sensor data, which has numerous applications in healthcare, sports, security, and human-computer interaction. Despite significant advances in HAR, critical challenges still exist. Game theory has emerged as a promising solution to address these challenges in machine learning problems including HAR. However, there is a lack of research work on applying game theory solutions to the HAR problems. This review paper explores the potential of game theory as a solution for HAR tasks, and bridges the gap between game theory and HAR research work by suggesting novel game-theoretic approaches for HAR problems. The contributions of this work include exploring how game theory can improve the accuracy and robustness of HAR models, investigating how game-theoretic concepts can optimize recognition algorithms, and discussing the game-theoretic approaches against the existing HAR methods. The objective is to provide insights into the potential of game theory as a solution for sensor-based HAR, and contribute to develop a more accurate and efficient recognition system in the future research directions.

## 1. Introduction

The present study is to investigate the new Human Activity Recognition (HAR) automatically identifies human activities using video images (CCV2 and RGBD) [1] [2], environmental sensors (WiFi, Radar, RIFD, WSN, PIR [3] [4] [5] or motion sensors (accelerometer, gyroscope, magnetometer) [6] that can be integrated into wearable devices or smartphones, which allows for continuous monitoring of human activities. HAR is also a significant research area with numerous applications in fields such as healthcare [1], sports, security, and human-computer interaction [2] [3]. HAR involves using machine learning algorithms to recognize and understand human behavior, which can provide valuable insights into individual and group activities.

The field of HAR has seen significant advances in the recent years, driven by the increasing availability of sensor data and the development of machine learning algorithms. One of the critical trends in HAR is the use of deep learning algorithms, which have shown promising results in recognizing and classifying human activities. Deep learning algorithms can automatically learn features from sensor data, reducing the need for manual feature engineering, and improving the accuracy of HAR systems.

However, they face several critical challenges common to machine learning and deep learning problems. According to the review paper [7], these challenges include feature extraction, annotation scarcity, class imbalance, distribution discrepancy, complex activities, data segmentation, computational cost, privacy, and interpretability.

In the following items, we will briefly address these critical challenges as they are highly relevant to the scope of our paper:

- Firstly, feature extraction is a critical issue in HAR, as it involves identifying and selecting the most relevant features from sensor data to recognize human activities accurately. Without effective feature extraction techniques, the performance of HAR algorithms will be limited. Therefore, addressing this challenge is essential for improving the accuracy and robustness of HAR systems.
- The scarcity of annotated data limits the ability to train and evaluate HAR algorithms. As a result, there is a need for more comprehensive and diverse annotated datasets to improve the performance of HAR systems.
- Class imbalance and distribution discrepancy are also critical challenges, as they can result in biased and inaccurate models. Addressing these challenges is critical for ensuring that HAR systems are robust and reliable, and can be applied to various scenarios and populations.
- Complex activities such as those involving multiple actions or interactions between people present a particular challenge for HAR. These activities are difficult to recognize, and require more advanced algorithms and processing techniques.
- Data segmentation is also an important challenge in HAR, as it involves breaking down continuous activity data into discrete segments for analysis. Developing effective segmentation techniques is critical for accurately analyzing activity data and detecting changes in behavior over time.
- Computational cost is another challenge in HAR, as processing large volumes of sensor data can be computationally intensive and time-consuming. Addressing this challenge is essential for improving the scalability and practicality of HAR systems.
- Privacy and interoperability are critical challenges, as they ensure that HAR systems are secure and compatible with other technologies and systems. Addressing these challenges is critical for ensuring HAR systems can be deployed effectively and safely in various contexts.

Game theory has emerged as a promising solution to address the mentioned challenges. Game theory is a mathematical framework that models interactions between multiple entities in various fields including economics, political science, and computer science. In the recent years, game theory has been increasingly applied to machine learning problems as a potential solution to improve recognition accuracy and optimize sensor configurations and recognition algorithms.

As the HAR is a classification task, it has many shared challenges with machine learning. Thus applying game theoretic approaches is an excellent idea to solve HAR problems [8]. Generally, the optimization or decision-making problems could be modeled by game theory based on mathematical forms [9]. They can find optimal strategies for learning agents with incomplete information about each other. In the recent years, many machines learning-based researches have been conducted using game theoretic solutions, for example, noise filtering [10], feature selection [11][12], heterogeneous data, and tagging unlabelled data [13].

This paper aims to provide insights into the potential of game theory as a solution for sensor-based HAR, and contribute to develop a more accurate and efficient recognition system. In 2016, Rekha *et al*. [8] presented a review paper on game theory and how it can be applied to machine learning. However, they were addressed to any HAR challenges. Tanmoy *et al*. also [14] highlighted the potential benefits of incorporating game-theoretic concepts into deep learning models, mainly focusing on the popular Generative Adversarial Network (GAN) architecture. The review discussed how GANs, essentially two-player zero-sum games, have successfully solved complex computer vision problems. This paper also provided an overview of the various real-time applications of game-theoretic deep learning models and valuable datasets in the field. One potential limitation of the paper [14] is that it may not cover many challenges in HAR. While the paper provides insights into how game-theoretic concepts can improve deep learning models, it focuses on computer vision problems and using GANs. It may not supply a comprehensive overview of HAR challenges such as sensor data collection, pre-processing, and model interpretability. Therefore, the researchers interested in applying game-theoretic approaches to HAR may need to look for additional literature to address these challenges.

As per the current literature, there is a lack of research in applying game theory solutions to the HAR problems. Therefore, our motivation is to explore the potential of game theory as a solution for HAR tasks. Our contribution will be to bridge the gap between game theory and HAR research by developing novel game-theoretic approaches for HAR problems. We aim to investigate how game theory can improve the accuracy and robustness of HAR models and how it can help address shared challenges in the field such as data scarcity, model interpretability, and real-time performance. Our goal is to demonstrate the potential of game theory in HAR and provide new insights into its applicability in other related research areas. Overall, our crucial contribution will expand the scope of research in HAR by introducing game-theoretic concepts and solutions to the field of HAR as following items:

- We introduce game theory as a novel approach to improve recognition accuracy and optimize sensor configurations and recognition algorithms in sensor based HAR.
- We review the previous work on sensor based HAR, highlighting the limitations of traditional approaches and the potential advantages of using game theory.
- We describe the different approaches to use game theory in sensor-based HAR including non-cooperative and cooperative games. It discusses the advantages and disadvantages of each approach and provides insights into how they can be used to improve recognition accuracy.
- We propose new approaches for using game theory in sensor-based HAR and evaluate their effectiveness compared to traditional approaches. It contributes to the development of more accurate and efficient recognition systems.
- We discuss the limitations of current game theory approaches, and suggest future research directions. It helps to guide the researchers in the field toward developing more effective and efficient recognition systems.

This study is divided into six comprehensive sections. Section 1 introduces Human Activity Recognition (HAR) systems and outlines our research motivations. Section 2 delves into sensor-based HAR and its challenges. Section 3 provides an overview of game theory and its applications in AI. Section 4 revisits HAR challenges and discusses game theory solutions. Section 5 proposes new research directions for sensor-based HAR using game theory. Finally, Section 6 presents the conclusion. The overall paper structure also can be visualized in Figure 1.

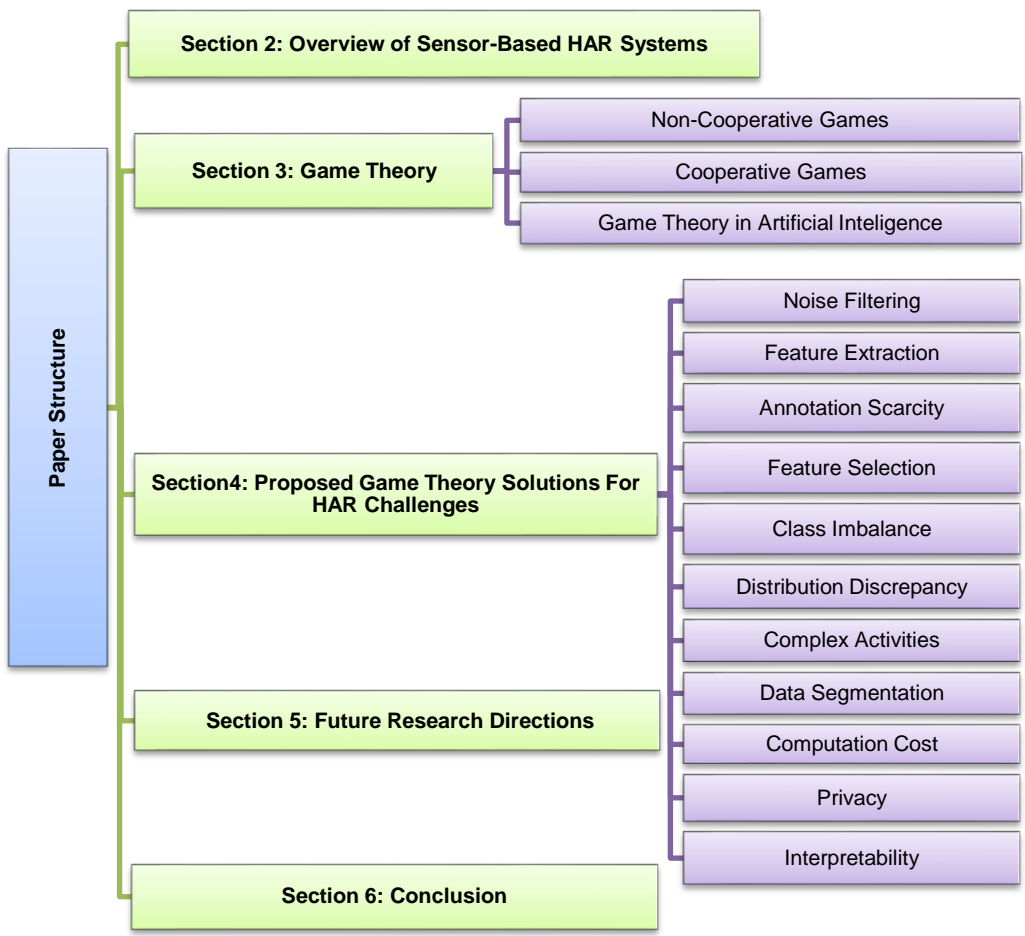

**Figure 1. Overall view of paper.**

## 2. Sensor-Based Human Activity Recognition Systems

Recently, HAR has become a popular research trend due to the easy accessibility of motion sensors in mobile phones or smartwatches, their low computation cost on portable devices with machine learning and deep learning developments in smart devices with powerful resources [15].

feature extraction, and classification steps. Most HAR systems are followed as the proposed architecture in Figure 2, inspired by [33] and [34]. In pre-processing section, first, noisy data is removed from raw motion data by noise filtering algorithms, then all raw data is converted to fixed lengths of windows data by a data segmentation mechanism.

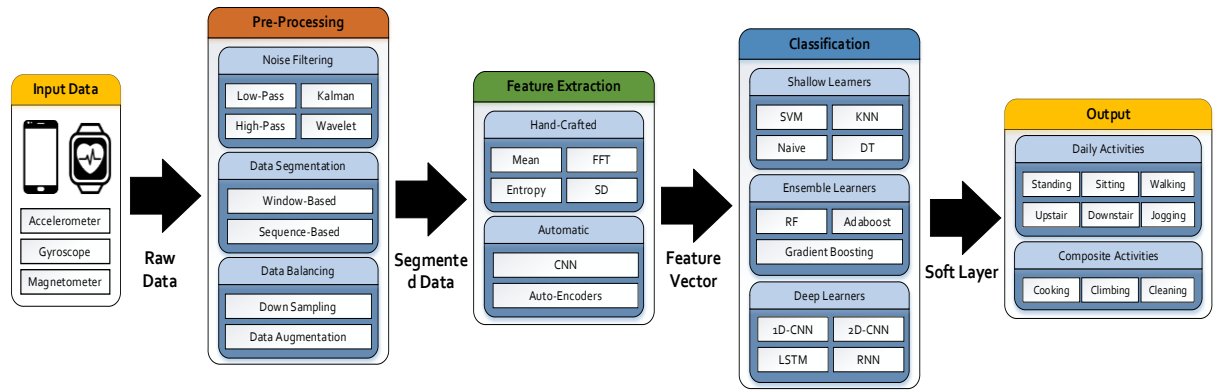

**Figure 2. General architecture of a sensor-based HAR system.**

Before that, the researchers used video-based sensors to monitor different user activities for simple and complex. However, they faced various fundamental challenges such as large video image datasets, increased computation costs, and user privacy problems [16]. Therefore, sensor-based HAR models are valuable research motives due to their cost-effectiveness, availability, variety, and accessible data collection. The accelerometer, gyroscope, and magnetometer are extensively used sensors embedded in smart devices (e.g. smartphones or smartwatches) and have been used in many studies.

In the recent years, HAR entered different research areas from a practical respective and played an influential role in various fields such as:

- Healthcare [17] [18] [19], treatment prevention [20] or fall detection [21]
- Smart home [22] [23], smart city [24] or energy-saving [25]
- Security areas such as user authentication systems [26]
- Sports [27]
- Interactive games [28]
- Cooking [29] [30] [31]
- Ontology [32]

In the mentioned research fields, simple and complex activity recognition as a classification task leads to direct and indirect objective realization and improves human life quality. It can be modeled by a standard architecture containing pre-processing,

Afterward, the fixed-window data balances in different classes. Next, feature sets are extracted from raw data through hand-crafted or deep-learning methods. Finally, different shallow learning models recognize simple or complex activities from extracted features.

In the proposed architecture, we supposed all data are labeled accurately as the ground truth labels in the data annotation procedure.

However, the data annotation procedure is time-consuming, tedious, and costly, and considers a fundamental challenge in HAR. They convert the datasets turned to unlabelled or inaccurate samples. Besides, the heterogeneity between training and test datasets is another challenge for HAR due to users' biological changes over time because of their dependency on environmental parameters. The researchers have recently employed semi-supervised learning methods such as co-training, active and reinforcement learning to solve data annotation and heterogeneous data.

A wide range of research and review papers has already been conducted in the HAR research field. In this regard, Demrozi *et al*. [35] comprehensively reviewed all studies from 2015 to 2019, and presented a survey paper for HAR. Nweke *et al*. [36] also presented a review paper based on motion sensors for HAR systems. They introduced popular essays and research challenges including 1) the usage of deep learning models in real-time, 2) the evaluation of pre-processing methods and how to tune the learning models' parameters, 3) the study of deep learning methods with large-volume data, 4) implementation of transfer learning systems for mobile phones, 5) developing the decision fusion-based deep learning methods for mobile and wearable devices, 6) solving imbalanced data problems, and 7) producing augmented data. Furthermore, Wang *et al*. [34] studied on deep learning-based activity recognition systems. They presented the following challenges: 1) identifying online activities using in-depth mobile learning, 2) recognizing more accurate unsupervised activities, 3) identifying high-level activity with flexible models, and 4) presenting lightweight models based on deep learning.

that the game theory analytical tools were not used in the HAR problems. Of course, the game theory proposes many optimization solutions for general problems related to machine learning but not specifically for HAR systems. We believe that game theory can assist HAR challenges with its different analytical tools, which will be discuss in how game theory can be applied to artificial intelligence and its sub-research fields in the following sections. In the present section, a brief introduction of applications and the relevant challenges of HAR are reviewed, and we will discuss the basic concepts of game theory and its applications in artificial intelligence in the following section.

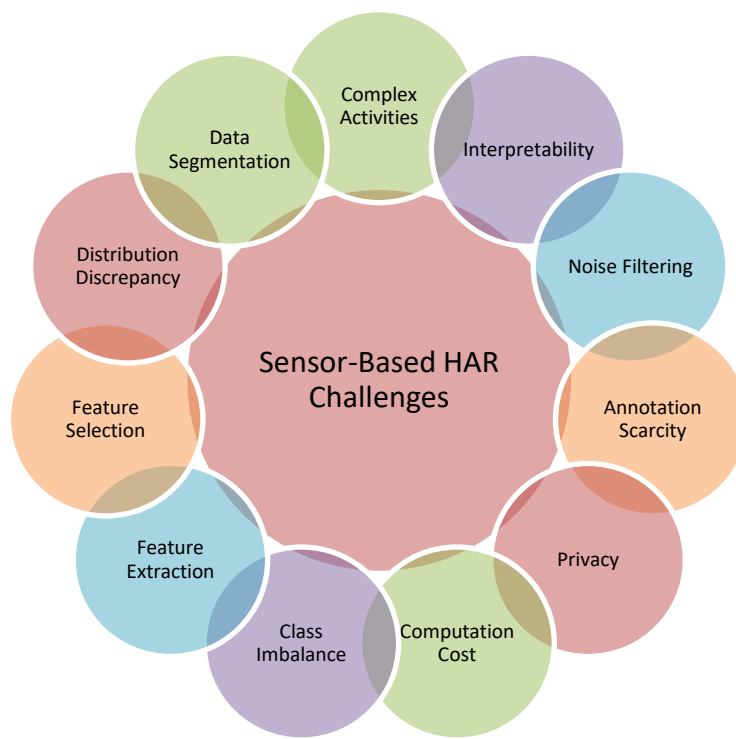

**Figure 3. Main challenges in sensors-based HAR systems** [7]

Moreover, Chen *et al*. [7] analyzed HAR systems' fundamental challenges and opportunities based on deep learning methods. Inspired by represented challenges [7], we study and analyze them individually and suggest analytical game theory tools for each challenge. Figure 3 presents our main HAR challenges, which we will discuss in Section 4. Based on the paper [7], the diagram refers to unsolved essential problems and should be addressed more closely because they provide proper research directions. For example, feature extraction, data heterogeneity, identification of complex activities, and interpretability of deep learning models are among the essential topics indicated with more sub-parts. As discussed earlier in the research motivations (Section 1), we found

**3. Game Theory**

The game theory analyses the different games with rationality and intelligent beings using mathematical models [9] and is utilized in social sciences, economics, biology, engineering, political science, international relations, computer science, philosophy, etc. In addition, it studies mathematical models in terms of strategy in which the success of one agent in selecting its actions depends on the others' choice. First, the game theory was developed in economics to perceive various economic behaviors such as companies, markets, and consumers. Then they developed moral or normative behavior theories, which scientists applied to rational behaviors [37]. Different game types (e.g. zero-sum / non-zero-sum, symmetric / asymmetric, simultaneous / sequential, cooperative / non-cooperative etc.) can be presented in the concept of game theory. One of the popular approaches presents three mathematical models in the following categories: 1) The strategic forms for non-cooperative, 2) the coalition forms for cooperative games, and 3) the extensive forms that apply in dynamic situations in computer science. This paper discusses strategic and coalition forms due to their applications to the related works of machine learning problems.

### 3.1. Non-cooperative games

A non-cooperative game is based on strategic decisions with $N > 1$ players with complete or partial information about the rules and all possibilities. Each player has its utility function, and we define the game as the following steps:

1. Number of $N > 1$ players as $P = \{P_1, P_2, P_3, \ldots, P_n\}$.
2. The action set for each player $P_i$ as $A = \{a_1, a_2, a_3, \ldots, a_n\}$
3. A strategy set $S_i$ assigned to the players from $S = \{S_1, S_2, S_3, \ldots, S_n\}$
4. A utility function as the payoff for each player is given by:
   $Utility_i = (a_1, a_2, a_3, \ldots, a_n) = i's$ payoff if the action profile is $(a_1, a_2, a_3, \ldots, a_n)$

Two or more players determine with known payoffs or quantifiable consequences to get the best outcome, and the players have a complete set of actions as their strategies perform according to possible situations in the game.

All participants receive a reward from a particular outcome as their utilities, known as the payoff in game theory. The players reach a common point where they have made their decisions and achieved the desired outcome as the solution. In solutions concepts, Nash equilibrium is the most famous solution with the game of $G = (N, (S_n)_{n \in N}, (u_k)_{n \in N})$ in strategy profile $s^{NE} = (s_1^{NE}, \ldots, s_n^{NE}) = (S_n^{NE}, S_{-n}^{NE})$ that is given by [38]:

$$\forall n \in N, \forall s_n \in S_n, u_n\left(S_n^{NE}, S_{-n}^{NE}\right) \geq u_n\left(S_n^{NE}, S_{-n}^{NE}\right) \quad (1)$$

In addition, the extended version of the above definition is called mixed strategy, which may be a vector of actions and their probability distributions. In mixed strategy, $\pi_n(s_n) \in (S_n)$ is a distribution that assigns a probability of $\pi_n(s_n)$ to each strategy of $s_n$ for each player $P_i$ in the game $G$ that is a profile $\pi^{NE} = (\pi_1^{NE}, \ldots, \pi_n^{NE}) = (\pi_n^{NE}, \pi_{-n}^{NE})$ such that:

$$\forall n \in N, \forall \pi_n \in (S_n), \tilde{u}_n\left(\pi_n^{NE}, \pi_{-n}^{NE}\right) \geq \tilde{u}_n\left(\pi_n^{NE}, \pi_{-n}^{NE}\right) \quad (2)$$

$$\tilde{u}_n(\pi_n, \pi_{-n}) = E(u_n) = \sum_{s \in S}\left(\prod_{j \in N} \pi_j(s_j)\right) u_n(s) \quad (3)$$

where $\tilde{u}_n$ is the expected utility for each player when they select the mixed strategy of $\pi_n$. Other suggested solutions are minimax as a zero-sum game, one of the most common non-cooperative and strategic form games with two players. A zero-game is represented as $N = 2$ players and $u_1(s1, s2) + u_2(s1, s2) = 0$ that one player tries to maximize his profit, as opposed to another trying to minimize his loss. Minimax is a strategy of always minimizing the maximum possible loss value.

### 3.2. Cooperative games

As discussed in non-cooperative and zero-sum games, strategic forms were focused on the set of choices of individual players, and each player would reach their goals. However, the cooperative games' nature was based on the players' ability to communicate in groups to improve their state and positions. In a cooperative game, we deal with players as coalitions that act harmoniously.

In each coalition, the players may choose the optimal strategies similar to strategic forms, but the goal is to analyze the possibility of communication between the participants. The coalition forms divide into the transferable utility (TU) and no transferable utility (NTU) games [39]. In the TU games, agreements are made between a subset of players, transferable between players. Nevertheless, the gain may not be transferable in NTU games. The coalition-based games define as a function $V$ as the maximum outcome that assigns every coalition $S$ to set $(S) \subset R^S$, which describes the feasible payoff for profiles. In TU-games, $V$ is a real-valued function if:

$$v = (v(S))_{S \subseteq C}, V(S) = \{u^S \in R^S : \sum_i u_i^S \leq v(S)\} \quad (4)$$

In addition, the solution concepts in coalition-based games are classified into set-valued and single-valued solutions, in which the first focus on stability, and the second is tailored toward fairness. The most popular methods are 1) *Core*, 2) $\varepsilon$ – *Core*, and 3) shapley value that is based on single-valued solutions [38]. Moreover, much extensive research has been conducted in parallel with algorithm design and game theory in the recent years, which has led to the emergence of a new field called algorithmic game theory [40]. Some researchers achieved significant developments in algorithmic game theory and artificial intelligence that looks different but have a strong relationship. Artificial intelligence problems encounter challenges solved efficiently using algorithmic game theory [8]. An overview of the proposed game theory solutions based on artificial intelligence methods is presented in Figure 4.

algorithms to consider the trade-off between coverage quality and energy consumption. Besides, Georgia *et al*. [44] developed a game-theory-based collective classifier framework that used the weighted majority voting system with game theory to identify the best classification model for the system input data. Their systems were based on artificial intelligence tools that emerged as single-agent or multi-agent systems and performed a specific job in a environment with collaboration. In multi-agent systems as sub-fields of artificial intelligence, the performance is defined according to the embedded rules in the design step; however, making a specific protocol is not simple for self-interested agents. By designing a mechanism, agents should be encouraged to agree to perform a utilitarian task.

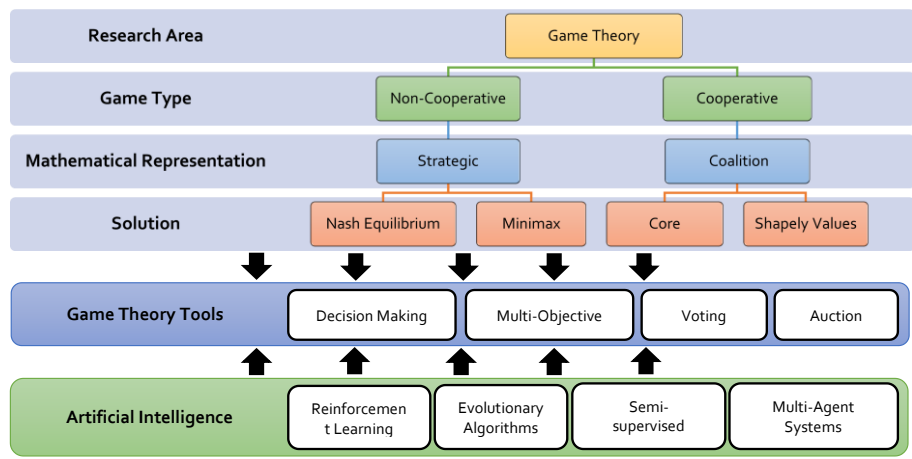

**Figure 4. General overview of the game theory research area in cooperative/non-cooperative types and its relations with artificial intelligence.**

### 3.3. Game theory in artificial intelligence

Artificial intelligence is efficient in many research fields such as robotics, business, education, economics, manufacturing, security, and data interpretation. They are typically developed with machine learning and deep learning methods that address challenges including feature engineering, computation cost, memory management, pre-processing, and hyperparameter optimization. The game theory recommends practical solutions to the challenges mentioned in decision-making-based systems. In this regard, the challenges are first converted into multi-objective or distributed optimization problems; then game theory-based mathematical models are applied to formulate them. For example, Shahrokhzadeh *et al*. [41] proposed a new distributed game-theoretic approach for Visual Sensor Networks (VSN) [42] [43] to cover the quality by disabling some switches and adjusting the sensors' orientation. They formulated the VSNs with the potential games and distributed payoff-based learning

Different game theory methods develop the mechanisms with the auction, debate, voting, bargaining, and negotiation approaches to satisfy the system's specific capabilities. English auction, Dutch auction, sealed first-price auction, Vickery auction, ideal auction, and zeuthen strategy are the most popular methods [45]. On the other hand, the Reinforcement learning method and other artificial intelligence sub-fields are introduced to enable agents to learn in an interactive environment using reward feedback from their actions and experiences. Reinforcement learning employs an input-to-output mapping as supervised learning and calls different action sets to perform reward and punishment tasks as positive and negative behaviors. Reinforcement learning can also be used in multi-agent systems, called multi-agent reinforcement learning (MARL). They guarantee the maximum convergence of learning algorithms to optimal Nash equilibrium. According to the game settings, the MARL approaches are divided into stateless games, team Markov games, and general Markov games. Each group helps multi-agent-based reinforcement learning systems regarding application and required information from the environment [46]. On the other hand, game theory has also contributed to deep learning problems in the recent years. They improve the results in deep learning models; in other words, they form the essence of deep learning models. In this regard, Hazra *et al*. [14] published a review paper, in which an extensive account of significant contributions took place in deep learning using game-theoretic concepts.

game theory via suggested solutions for machine learning challenges, which are familiar with HAR challenges. Inspired by this communication, we introduce new HAR challenges based on game theory in the following sections.

As shown in Figure 5, a relational diagram is proposed using game theory analytical tools on shared machine learning challenges, which can be applied to HAR systems. At the center of the diagram, we have the HAR challenges that game theory analytical tools tend to help to improve the recognition performance or model efficiency in HAR systems.

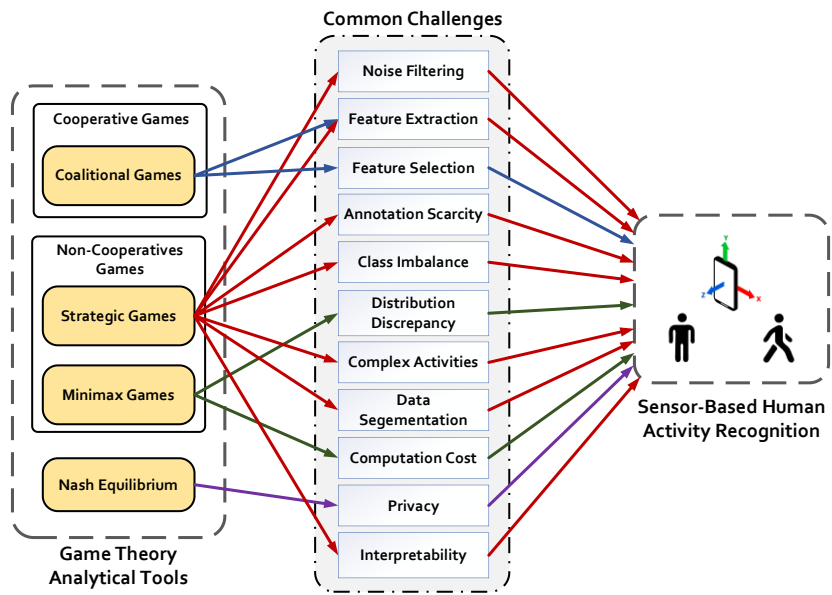

**Figure 5. A rational diagram depicting how game theory can help to solve challenges in Human Activity Recognition (HAR) that are common with both machine learning and game theory.**

Another sub-field of artificial intelligence is evolutionary algorithms that suggest several optimization algorithms to find the best profits. Game theory analytical tools can be used in evolutionary algorithms to find optimal choices. Sohrabi and Azgomi [47] prepared a review paper about using game theory for optimization algorithms to improve the performance of evolutionary-based systems.

**4. Proposed Game Theory Solutions to Sensor-based HAR Challenges**

In Section 2, we identified 11 main challenges of HAR systems according to [7]. On the other hand, literature was reviewed on cooperative and non-cooperative games for artificial intelligence issues in Section 3.

Given that HAR systems have many shared challenges with machine learning problems, an indirect relationship is found between HAR and

The literature review identified the relationships between the application of game theory in shared machine learning challenges, and we will discuss how to use them for HAR challenges in the following sub-sections. We first review related works of HAR challenges, then the game-theory-based methods are proposed for each challenge. It is to be noted that suggested methods with game theory have been applied to general machine learning problems, which we later plan to develop for HAR challenges in future works.

**4.1. Noise filtering**

Many different sensors such as accelerometers, gyroscopes, and magnetometers are utilized as raw signals for input data in HAR systems, and they probably contain electronic noise due to the sampling error. The samples with noisy data are converted into fixed time-length windows using various segmentation methods for training or testing the HAR learning models. Noise can negatively affect the performance of the activity recognition system; therefore, pre-processing is performed on the raw data using filtering methods that eliminate the existing noise, and then use the data. Atnar *et al.* [48] compared the efficiency of Wavelet, Kalman, Low-Pass, and Median as popular filtering algorithms in HAR systems. Accordingly, the low-pass method had the best performance in real-time systems; significant waveform delay depends on filter order cut-off frequency, lowest performance with waveform delay for signal-to-noise ratio (SNR), and mutually highest performance without waveform delay.

Darwish *et al.* [10] proposed a noise reduction approach based on non-cooperative games on a cognitive radio network that found the best noise reduction algorithm. Each de-noising method was represented as a player with pre-defined strategies from adjusting affective variables. Then, a payoff outcome is obtained from the players' objective functions. The system is based on a non-collaborative game with the standard form that is given by:

$$G = N, A, \{u_i\} \qquad (5)$$

where the variable $N$ is a set of decision-makers (players), $A$ is the Cartesian product of possible actions $(A_1 \times A_2 \times ... \times A_n)$ for player $i$, and $\{u_i\}$ was a set of utility functions of each player as the outcome of the players. Therefore, the outcome of the actions (for each $A$) is selected by $a_i$ for the player $i$, and $a_{-i}$ from all players' chosen actions. $a_i$ and $a_{-i}$ are an exclusive choice of actions that Nash equilibria can specify by each player as the action tuples.

One possible approach for noise reduction in sensor-based HAR using non-cooperative games is to model the interaction between the sensor data and the noise as a non-cooperative game between two players: the sensor data and the noise. In this game, the sensor data player aims to classify human activity correctly, while the noise player aims to introduce as much noise as possible to mislead the sensor data player.

Each player has a set of strategies corresponding to different actions they can take. For instance, the sensor data player's strategies could be different feature selection techniques, while the noise player's strategies could be different types and noise levels.

A Nash equilibrium could be computed to find the optimal solution, representing a set of strategies where neither player can improve their payoff by unilaterally changing their strategy. In other words, it is a stable solution where neither player has the incentive to deviate from their strategy. Once the Nash equilibrium is found, the sensor data player can use the corresponding strategy to classify the human activity while mitigating the noise introduced by the noise player. Using this approach, the sensor data player can improve its classification accuracy, even in noise.

To formulate the proposed solution, given a set of sensor readings for HAR, the goal is to develop a noise reduction approach based on non-cooperative games. Let $X$ be the set of sensor readings and $Y$ be the set of activities. Let $x_i = \left(x_i^1, x_i^2, ..., x_i^n\right)$ be a vector of n sensor readings for activity $y_i$ where $i = 1, 2, 3, ..., m$ are the total number of activities. Let $z_i = \left(z_i^1, z_i^2, ..., z_i^n\right)$ be a vector of n noise readings for activity $y_i$. The observed sensor readings can be represented as follows:

$$u_i = x_i + z_i \qquad (6)$$

where $u_i$ is the vector of observed sensor readings for activity $y_i$. The goal is to estimate $x_i$ from $u_i$ by formulating a non-cooperative game among the sensors. Let $S = \{S_1, S_2, ..., S_n\}$ be the set of sensors, where $S_n$ represents the *n*-th sensor. Let $f_i = (S_i, x_i)$ be the payoff function for the *i*-th sensor, representing the quality of the sensor reading for activity j. The goal is to find a Nash equilibrium among the sensors to estimate $x_i$ from $u_i$.

However, it is worth noting that this approach has some limitations. One of the main limitations is that it assumes that the noise player's strategy is known, which may not always be the case in practice.

Finding the Nash equilibrium can also be computationally expensive, especially for large datasets and complex models. Therefore, further research is needed to investigate the effectiveness and scalability of this approach for real-world HAR applications.

### 4.2. Feature extraction
Feature extraction is a vital component of the HAR systems, which extracts a robust feature set from raw input data to discriminate various activities. Temporal feature extraction is one of the most popular types performed manually (hand-crafted) or automatically using end-to-end deep learning models. The hand-crafted approaches transform the raw data into feature vectors with statistical calculations in the time or frequency domains.

As shown in Figure 6, hand-crafted approaches reduce input data due to initial pre-processing steps, leading to a real-time HAR system [49]. Despite the outstanding advantages of hand-crafted feature extraction, they have some fundamental problems that lead to low performance with coarse labels and a lack of scalability. Therefore, despite the outstanding advantages of hand-crafted feature extraction, some fundamental problems are low performance with coarse labels and a lack of scalability. Therefore, end-to-end deep learning methods are proposed as advanced approaches for extracting more efficient features. In the initial layers of these methods (convolutional layers), a set of robust features are automatically detected by the machine from raw data, and then these features are referred to as the hidden layers of the learning network.

window length is a hyper-parameter in the learning models that must be adjusted in the optimization stage.

In addition to RNN-based methods, LSTM deep learning models are used as two-way or ensembles, and various research projects have been conducted on popular datasets and provided good results [21] [52]. Unlike RNNs and LSTMs, CNN models do not need stream segmentation and use a small, fixed-length kernel on each vector input direction [53] [54]. Different motion sensors like accelerometers and gyroscopes can employ parallels to improve the performance of HAR systems. These feature extraction models are known as multi-modal feature extraction, where diverse data forms are generated by integrating the sensors in various cases. The feature fusion and ensemble classifier are applied to the multi-modal problem to integrate the data. Münzer *et al*. [55] presented different solutions in four categories that led to the best combination of sensors using specific policies related to Feature Fusion.

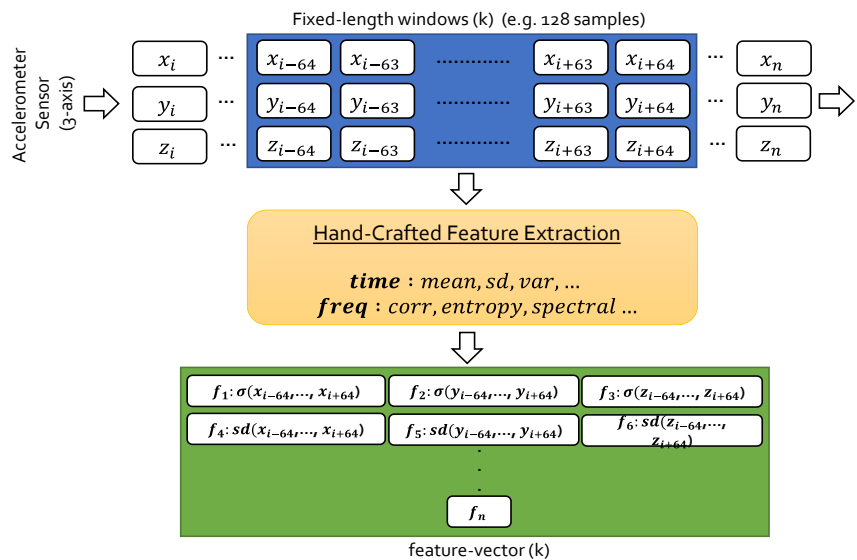

**Figure 6. Procedure of the feature extraction task is based on the hand-crafted approach in HAR systems.**

They can also be used for deep learning-based HAR systems, converted from time series to frequency domains, then processed by a 2D-CNN layer [50]. Alternatively, the raw samples or their segmentation are applied with a fixed length in specific time windows and entered into deep learning systems as train or test samples. Many studies have been presented based on the literature's RNN, LSTM, and CNN methods.

Due to the stream form of input data, the segmentation is first utilized (all the different motion sensors are placed together in one segment), and then the performed segments are separately entered into an RNN cell [51]. The

The Ensemble Classifier strategy identifies the best combination of sensors in another space based on the ensemble learning attitude and data participation in different modalities [56]. In addition to temporal and multi-modal feature extraction methods, statistical feature extraction is another model in the deep learning space that extracts significant features based on data engineering in statistical information.

In this regard, Qian *et al*. [57] identified user activities in a Distribution-embedded Deep Neural Network (DDNN) by performing and integrating statistical features. Table 1 summarizes recent works on the main three feature extraction approaches in HAR systems and their advantages and disadvantages. According to the literature review from Table 1, the decision-makers can apply the game theory to provide feature fusion or ensemble classifier strategies for HAR applications.

In the proposed equilibrium, the aim was to minimize the expected payoff for each player. In similar studies, Bruce *et al.* [59] also applied game theory for automated analysis to fuse the hyperspectral imagery data with remotely sensed data into health monitoring systems for precision agricultural applications.

**Table 1. Comparison of various Feature Extraction Approaches in sensors-based activity recognition systems** [7].

| Approach | Related works | Advantages | Disadvantages |
|---|---|---|---|
| Temporal | [49] [50] [51] [52] [53] [54] [109] [110] | Simple | Limited to coarse activities |
| | | Real-time performance (hand-crafted) | High computation cost (deep models) |
| | | Extract local and global temporal features (deep models) | Experience-dependent |
| | | | Low stability |
| Multi-Modal | [55] [56] [111] [112] [113] | Simple | Limited to coarse activities |
| | | High stability | Unstable performance |
| | | Extract hierarchical features | Complex structure |
| Statistical | [57] | Good Interpretability | Dependent on domain knowledge |

Similarly, Yazidi *et al.* [58] categorized the input sensors as reliable and unreliable with a repeated strategic game and set of learning automata (LA). The problem was formulated in an $N$ population sensors $S = \{s_1, s_2, ...., s_N\}$, that the output of each sensor $s_i$ was referred to as $x_i$. The actions of sensor $s_i$ was denoted by $a_i$ and $a_{-i}$ to other player's profile. Also the reward of each player was calculated by the following equation:

$$r_i(t) = \begin{cases} 1, & if\ |G_{a_i}| = 1 \\ \dfrac{\sum_{k \in G_{a_i}, k \neq i} I\{x_k(t) = x_i(t)\}}{|G_{a_i} \setminus \{i\}|}, & otherwise \end{cases} \quad (7)$$

where $G_{a_i}(t)$ was denoted the set of the same actions for the sensor $s_i$ at instant time of $t$. It assigned 1 to $r_i(t)$ when $s_i$ was the only sensor in $G_{a_i}$ and also $|G_{a_i}(t)| > 1$ for the number of players who agreed with sensor $i$ across those sensors that have selected the same group as $s_i$ at time $t$. In addition, the authors [58] defined the utility function of player $i$ by:

$$u_i(a_i, a_{-i}) = E[r_i | (a_i, a_{-i})] \quad (8)$$

that was the expected payoff when a pure strategy $a_i$ for had been selected, while others chose the profile $a_{-i}$.

The authors analyzed the coalition and strategic games on feature-level and feature-decision fusion tasks (as feature grouping problems) for hyper-temporal-hyperspectral and reported a satisfactory result compared to other studies. They applied the game theory analytical tools to multi-classifier decision fusion systems in different stages. The input data was partitioned into $N$ sub-datasets, and each was classified independently. Besides, a Nash equilibrium was used to find a steady-state solution to the feature grouping problem to maximize the payoff for all players or feature groups.

In HAR, different motion sensors (e.g. accelerometer and gyroscope) can be fused, improving the model performance. However, choosing the best combination of sensors is a challenge, determined by a decision-making task or optimization problem. As the feature or sensor fusion recommended for multi-modal sensor problems, the idea of repeated strategic games and automata learning (LA) could be applied, inspired by [58].

To formulate the possible solution, let there be $m$ different sensor modalities, each providing a stream of observations. Let the $i$-th modality have a sampling rate of $f_i$ Hz, and let the observations from each modality be denoted by $X_i^t$ where $t$ is the discrete time index. The goal is to perform a HAR using these multi-modal sensor inputs.

The new method for sensor fusion involves combining the observations from all modalities at each time step $t$ into a single observation vector $x_t \in \mathbf{R}^n$, where $n$ is the total number of features in the fused observation vector. It can be achieved using a feature extraction step for each modality, followed by a concatenation operation to form the fused observation vector. Mathematically, we can write:

$$x_t = \left[ f_1(X_t^1), f_2(X_t^2), \ldots, f_m(X_t^m) \right]^T \quad (9)$$

where $f_i(.)$ is the feature extraction function for modality $i$.

Next, the fused observation vector $x_t$ is used as the input to a machine learning algorithm to perform HAR. This algorithm can be a repeated strategic game or automata learning (LA) algorithm, depending on the problem. Mathematically, we can write:

$$y_t = HAR(x_t) \quad (10)$$

where $y_t$ is the predicted human activity label at a time $x_t$.

### 4.3. Feature selection

Feature extraction of raw data from motion sensors produces many features in different time and frequency domains that may be inefficient. The inefficient features reduce the efficiency of activity recognition systems in the classification phase. Thus, proper feature selection methods lead to selecting a subset of features that improve the accuracy of recognition systems and increase system execution time. The feature selection algorithms determine the valuable features and the features that should be removed. A score is assigned depending on feature roles during learning, and inefficient features are eliminated based on this scoring system. ReliefF, Correlation-based Feature Selection (CFS), and Fast Correlation Based Filter (FCBF) are popular approaches to selecting the best features in the HAR field [60].

In related studies of feature selection, Cohel et al. [11] proposed a new feature selection method based on coalitional games that remove inefficient items with segmentation, dominance mechanism, and multi-perturbation shapely analysis (MSA) to estimate the usefulness of game theory solutions. Generally, the coalitional games are defined by a pair of $(N, v)$, where $N = \{1, \ldots, n\}$ is a set for all players and $v(S)$ are a real-value function to represent the payoff for every $S \subseteq N$. The game theory represents the contributions to each player based on the Shapley value that is given by:

$$\Phi_i(v) = \frac{1}{n!} \sum_{\pi \in \Pi} \Delta_i(S_i(\pi)) \quad (11)$$

where the $\Pi$ is a set of permutations over $N$, $S_i(\pi)$ was a set of players appearing before the $i$th player in permutation $\pi$. This method transformed into the feature selection challenge, where players have created the features of a dataset and resulted in a payoff from $v(S)$ measuring the performance of the classifiers obtained using the set of features $S$. In the proposed method, they successfully optimized the performance of the classification problem over unseen data such as accuracy, balanced error rate, and area under the receiver-operator-characteristic (ROC) curve. In other studies, Chu *et al*. [61] proposed a new method of selecting greedily collective features based on Shapely values, in which all the features are scored and selected as the best subset.

Game theory also is applied to feature selection for HAR for the first time by Wang *et al*. [62]. They introduced an ensemble empirical mode decomposition (EEMD) based on features and a feature selection system based on shapely values to estimate each feature's weight, and then a subset of the candidate features was selected with the highest weight. Most feature selection methods ignore features that act as a group with discriminatory power but are weak as individuals. However, cooperative game theory-based feature selection methods were proposed to select optimum feature subsets robust to a special status [62]. Their experimental results show that the proposed feature selection method selected fewer features and provided higher accuracy than other popular methods.

### 4.4. Annotation Scarcity

HAR is typically based on supervised learning methods and requires correctly labeled data from proper data collection procedures. However, data collection in sensory data procedures faces many challenges, such as time-consuming data annotation and noise in labeled large datasets. Due to the mentioned problems, a few samples might only be labeled correctly, and some samples lack labels.

The unsupervised and semi-supervised learning methods could be suggested to solve the problem. Unsupervised learning methods are utilized to analyze unlabelled data; for instance, generative learning models such as Deep Belief Networks (DBNs) and Auto-Encoders have led to successful studies about unsupervised methods. They extract individuals' efficient characteristics and behavioral patterns from unlabelled data used in labeled data [63][64]. As observed in the literature, unsupervised methods successfully analyzed the miss-labeled data; however, they might not automatically assign the labels to their corrected classes. Thus, semi-supervised methods alternatively have been presented, such as co-training, active learning, and data augmentation. Co-training learners that act similarly to human learning establish an interaction between experience and knowledge to generate knowledge based on experiences. In HAR, the experience is the labeled data; other data is selected according to the developed models by the labeled data and added to the labeled collection [65]. Active learning is another popular method that analyses and tags unlabelled samples using external factors. Instead of using labeled samples as previous experiences, the system starts labeling the unlabeled sample of individuals [66] [67]. In addition to the described methods, generating artificial data can solve unlabelled data problems. The data augmentation generates the artificially labeled data from the ground-truth small data set, and they are combined with an authentic sample to train recognition systems. The Generative Adversarial Network (GAN) generates synthetic data and authentic samples. Wang *et al.* [68] conducted Sensory GAN research in the HAR field, which yielded satisfactory results in dealing with small data sets. As discussed earlier, the presented approaches to solving the Annotation Scarcity problem have many strengths and weaknesses. A list of advantages and disadvantages of different technologies is presented in Table 2.

perspective. They can be used as rules-based data labeling mechanisms to trade costs and quality on data augmentation optimization or to provide the optimal predictors for annotation scarcity problems. In this regard, Yang et al. [69] developed a cost-effective framework for rules-based data labeling based on a game-based crowdsourcing approach as CROWDGAME. The main idea was to generate high-quality rules to reduce costs and maintain quality. Inspired by the CROWDGAME, two players play a game with generator and refuser roles to identify high-quality and low-cost rules; a refuser was provided to express the reasons for rejecting data tuples for tagging by examining some tuples. In addition to the unlabelled data problem, overfitting and generalization operations reduced the accuracy of learning models. In the proposed method [69], a minimax strategy was introduced, and the efficient task selection algorithms were given by:

$$O^{R_q^*, E_q^*} = min_{R_q} \ max_{E_q} \ \Phi(R_q|E_q) \qquad (12)$$

where $R_q$ and $E_q$ respectively were denoted by the sets of rules and tuples that were selected by generator and refuser rules, $\Phi(R_q|E_q)$ was calculated to estimate the loss based on a set of $E_q$ tuples that are checked by crowds, and the optimal task sets of $R_q^*$ and $E_q^*$ CROWDGAME reaches it. In other studies, Behpour *et al.* [70] optimized the testing time efficiency and improved the accuracy of such problems by developing a game-theoretic interpretation of data augmentation for object detection. The image-based object recognition systems were presented, forcing the learning models to use the most rigid distributions of perturbed samples—this procedure aimed to find the optimal adversarial perturbation from ground-truth data to push the predictive models.

**Table 2. Advantages and Disadvantages of Different Approaches for Annotation Scarcity** [7]**.**

| Approach | Related works | Advantages | Disadvantages |
|---|---|---|---|
| Unsupervised | [63] [64] | Prepare a model without labeled data | High computation cost It depends on ground-truth data for activity learning |
| Semi-supervised Co-training | [65] | Benefits from labeled and unlabelled data Automatic label assigning mechanism to unlabelled data Independent to human labeling | Two data modalities in minimum case Multiple classifiers are required for each iteration |
| Semi-supervised Active-learning | [66] [67] | High-performance accuracy | Dependent on human labeling |
| Data augmentation | [68] | Prepare a generalized model | Using less unlabelled data |

The game theory has been applied in similar studies to solve the unlabelled data problem from a public

They proved that the game-theoretic solution in (Nash equilibrium) provides an optimal predictor and data augmentation distribution. According to the experiments, their proposed approach improves performance by 16%, 5%, and 2% for ImageNet, Pascal VOC, and MS-COCO detection.

Previous data augmentation approaches used less unlabelled data for HAR. As a result, applying rule-based data labeling ideas from [69] can improve the HAR models' performance facing weakly annotated datasets. We suggest a CROWDGAME-based data labeling system with two players as rule generator players for creating the answers rule validation tasks, and the rule refuser players reply to the tuple of the answer rules, where annotated label of a data tuple was corrected. On the other hand, the proposed method of [70] can also be a helpful idea for solving the annotation scarcity problem in HAR, which the following definition can define. Given a dataset of human activity data tuples, where a feature vector represents each tuple $x_i$ and an associated label $y_i$, the goal is to generate accurate labels for the entire dataset when annotated data are scarce. Expressly, we assume that only a limited number of annotated data tuples are available for training a supervised learning model, and we wish to augment the dataset with additional accurately labeled data tuples to improve the model's performance. In addition, the notation of the suggested model follows:

- Let $D = (x_i, y_i)_i = 1^N$ be the dataset of $N$ data tuples, where each $x_i$ is a feature vector, and $y_i$ is the corresponding label.
- Let $D_L = (x_i, y_i)_j = 1^L$ be the subset of $D$ that human annotators have labeled, where L << N.
- Let $D_U = D \, D_L$ be the subset of $D$ that has not been labeled.
- Let $f(x)$ be the true underlying function that maps feature vectors to labels.
- Let $h(x;w)$ be a supervised learning model that maps feature vectors to predicted labels, parameterized by $w$.

By leveraging the knowledge and expertise of multiple players, a CROWDGAME-based data labeling system enhances the accuracy and reliability of labeled data by generating additional accurately labeled data tuples.

### 4.5. Class imbalance

The classification of imbalanced data is an essential issue in HAR. The "imbalanced dataset" refers to a dataset in which the number of samples varies significantly in different classes. The classes with the minor sample are called minorities.

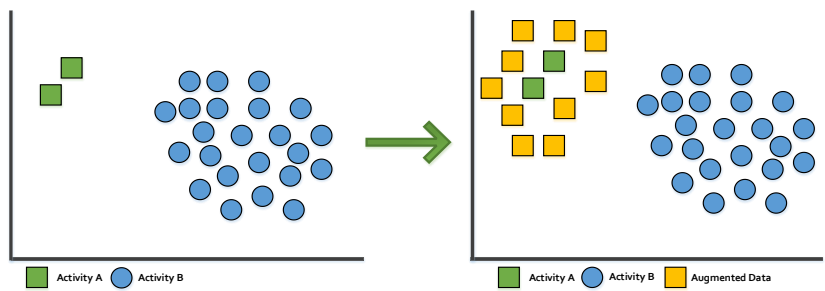

**Figure 7. Solving class imbalance with augmented data.**

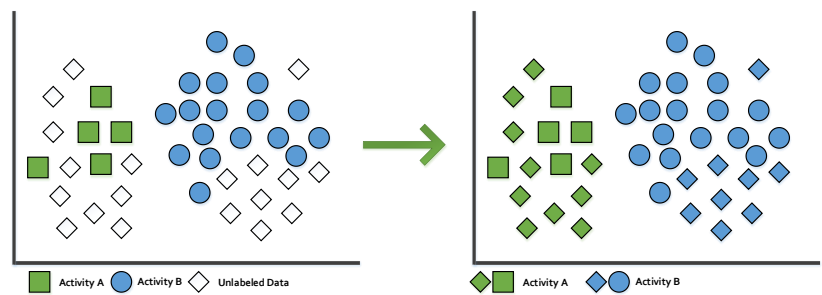

**Figure 8. Solving class imbalance with co-training.**

Since most learning methods train a classifier assuming an equal training sample in each class, the data are usually not balanced in different classes. Therefore, it causes inferior prediction because the training of the minority class is not adequately performed. The class imbalance problem is also expected in the HAR field since some daily activities occur more than others, such as walking activities. For example, in Figures 7 and 8, samples are the minority classes as a pure activity solved using various approaches including generating augmented data for low classes [54] or co-training to identify new unlabelled data samples to balance between the classes [65].

Almogahed *et al.* [13] designed a general oversampled filtering method with a non-cooperative game to solve the class imbalance problem. In this method, all samples consider the game players who interacted to obtain the dependent payoffs and estimate their membership probabilities alongside some synthetic data using the SMOTE (Synthetic Minority Over-Sampling Technique) algorithm. More precisely, the players are divided into $I_c$ class, which corresponds to minority and majority samples, $I_u$ shows the unlabelled and also synthetic samples. For each player $i$, the number of $b$ neighbors has computed all utility functions afterward for $i = 1, \ldots, I_u$ calculated for each player that has interaction with b neighbors of $j$ as given by:

$$u_i(x) = \sum_{j \in I_\emptyset \cap I_u} \left( x_i^T A_{ij} x_j \right) + \sum_{d=1}^{2} \sum_{j \in I_\emptyset \cap I_{c|d}} \left( x_i^T A_{ij} e_j^d \right) \quad (13)$$

where $d = 1$ and $d = 2$ define the minority and majority classes, respectively, the players of $I_u$ are in interacting with all players in $I_\emptyset$ and $A_{ij}$ as a partial matrix that computes the average payoff in the whole population. Furthermore, fitness is defined as $x_i^T A_{ij} e_j^d$ for $w_i(x)$, which is a utility function. As SMOTE is a well-known algorithm [71], using the proposed method from [13] can be helpful to class imbalance problems in HAR.

The proposed idea for HAR is to use a general oversampled filtering method with a non-cooperative game to overcome the class imbalance problem. The method involves using the SMOTE algorithm to generate synthetic samples from the minority class and modeling the interaction between players in the HAR dataset using game theory. This approach can improve the classification performance of the HAR system by providing a better understanding of player behavior.

**4.6. Distribution discrepancy**

The collected data using various motion sensors from multiple human activities is a distributed anomaly that affects learning models' performance in the HAR called the distribution discrepancy challenge. The HAR problems are presented in three groups: distribution discrepancy in users, time, and sensor data. In the first group, distribution discrepancy in data appears since each user has different patterns in their daily activities. In the second group, different activities will occur over time due to changes in patterns, and in the third group, there is a discrepancy in the distribution of data due to various factors such as location and the environment of sensor sampling. The use of transfer learning can be an appropriate method to solve many problems such as distribution discrepancy. For instance, regarding user-based methods, since some users are in training and others are in the testing phase, the accuracy of the learning model in the identification phase is reduced, and personalized learning models can be employed for a specific user. In this method, model personalization is performed using a small amount of testing phase data based on transfer learning or incremental learning, and the final model is tuned [72] [73].

Three fundamental problems including concept drift, concept evolution, and open-Set occur in the time distribution differences. A logical shift happens in training and testing samples over time for the concept drift, reducing classification accuracy. Continuous updating training models with new training samples can solve this problem [74]. In concept evolution, new activity emergence in the test phase leads to a similar problem, which is extremely important in the learning models. Using mid-level features such as raising and lowering the arms or lowering and raising the legs can solve this problem, and their combination leads to new activity identification [75]. Finally, as an up-to-date research trend, the open-set problem is the activity in a new space that cannot be included in the testing phase. GAN networks are used to solve the artificially generating a negative subset; therefore, it is only considered an anonymous activity [76].

However, factors such as instance sensors, type, position, environment, and layout can affect activity recognition accuracy in distribution with sensors. For instance, instance sensors occur when one in the test step has the same configuration as another sensor in the training phase or even with the exact specifications. Replacing a test phase sensor with a training phase sensor, despite both having similar configuration specifications, introduces an issue. Even when the sensor models and new devices share similar configurations, a slight alteration in the input data can be observed, which in turn, diminishes the overall system quality. Existing literature suggests that generating integrated samples via GANs can fortify models against the variations induced by sensor samples [77]. Moreover, changing the sensor type or mounted position on the user's body causes a difference in data distribution.

axial distribution discrepancy challenge and assist in the learning step on transfer learning.

In HAR, limited studies were proposed to find high-quality source data in neural network-based transfer learning. Developing a selective transfer learning method based on a minimax game can be an exciting opportunity for distribution discrepancy challenges in HAR.

Table 3. Main challenges for distribution discrepancy in HAR systems.

| Type | Challenges | Solutions | Related works |
|---|---|---|---|
| User | Distribution divergence between the training data and the test data | Personalized deep learning models | [72] [73] |
| | | Transfer learning | [114] [115] |
| Time | Concept drift problem | Incremental training | [74] [116] |
| | Concept evolution problem | Decompose activities into mid-level features | [75] |
| | Open-set problem | Generative adversarial networks | [76] |
| Sensor | Discrepancy between sensor instances | Data augmentation with GANs | [77] |
| | Different sensor types and positions | Transfer learning | [78] |
| | Sensor layouts and variety in environments | Generative adversarial networks | [77] |

In previous studies, transferring knowledge across different data domains was presented to solve the problem [78]. The sensor layout and environmental influences (due to obstacles) lead to distribution discrepancy in activity recognition systems based on device-free devices. Using a feature that leads to classifying different activities independent of the environmental parameter assists in solving these problems [79] [80]. In Table 3, we have summarized the main challenges of distribution discrepancy for HAR alongside the possible solutions for each challenge according to the literature.

The literature review presented in the following section indicates that taking advantage of transfer learning that can solve distribution discrepancies in HAR. On the other hand, many game-theory-based transfer learning methods were proposed for HAR; for example, Wang *et al.* [81] developed a transfer learning model to select the Mini-Max game's best data. In their proposed method, a two-axis mechanism is developed, including 1) a discriminator to maximize the differences between source and destination data and 2) a selector, as an attacker, to minimize the slightest difference between the source and destination data. The source data train the transferrable model and then, by giving a reward, leads the selector module to the optimal response. HAR can employ this mechanism to select the best source data in the

The possible formulation of the new method for selecting the best source data in the axial distribution discrepancy challenge that assists in the learning step on transfer learning for HAR with the mini-max game is as follows:

Let $S = \{S_1, S_1, \ldots S_n\}$ the set of candidate source datasets represent the $i$-th source dataset. Let $T$ be the target dataset, and let $F$ be a feature extractor model that maps input data to a high-level feature space. We aim to select the source dataset $S_i$ that is most relevant to the target dataset $T$ such that the feature distribution discrepancy between $S_i$ and $T$ is minimized.

We propose a new method called "Adversarial Source Selection (ASS)" that uses a mini-max game to select the best source dataset. The ASS method involves the following steps:

1) Initialize a target classifier model $G$ using the target dataset $T$ and a source feature extractor model $F$ using a randomly selected source dataset $S_i$.

2) Train a domain discriminator model $D$ to distinguish between the features extracted from the source $S_i$ and target datasets $T$. The discriminator $D$ is trained to maximize the source dataset's classification accuracy while minimizing the target dataset's classification accuracy $T$.

3) Train the source feature extractor model $F$ to generate features that fool the domain discriminator $D$, while minimizing the classification loss of the target classifier $G$.
4) Repeat steps 2 and 3 for all source datasets in S, and select the source dataset $S_i$ that yields the lowest feature distribution discrepancy between $S_i$ and $T$, as measured by the domain discriminator $D$.
5) Finally, fine-tune the target classifier $G$ using the selected source and target datasets $T$.

Using the ASS method, we can effectively select the most relevant source dataset for transfer learning in HAR, while minimizing the feature distribution discrepancy between the source and target datasets. This can lead to improved performance and generalization of the target classifier model.

**4.7. Complex activities**

According to the literature, most research related to HAR focused on simple activity recognition (such as walking, standing, and running) that does not require high computation costs. In addition to simple activity-based HAR systems, other HAR models are presented to identify more complex activities. Composite, concurrent, and multi-occupant are three of these complex activities. Table 4 lists different types of complex activities and proposes possible solutions for each type according to the related works.

alongside the LSTM network to identify concurrent activities for classifying high-level parallel forms activities from the output of the CNN networks. The concurrent activities are done simultaneously (such as talking on a cell phone and eating simultaneously); such activities are considered multi-label data performed by a user. Different motion and environmental sensors and several binary classifiers are applied in the parallel activity recognition system to check the accuracy of simple activity associative in the input data [83]. Authors in [84] proposed a better solution based on an LSTM multi-layer network, which identifies and evaluates each activity in the separate layers leading to parallel operations with recognition operations. Besides, A group of users usually performs multi-occupant activities concurrently or interactively. For example, eating and watching TV simultaneously is a parallel activity; playing tennis and football is interactive [85]. Also, different scenarios are proposed in designing multi-occupant activity recognition systems; however, only motion sensors are used separately on users in similar activities. Therefore, the system is considered a single-occupant activity; however, taking advantage of the environmental sensors in a parallel form causes a fundamental challenge in communicating with motion and environmental sensors that should be considered [86]. Concurrent models can be viewed as multi-label learning methods. In this light, Jung *et al*. [87] introduced an efficient ranking method for multi-label data tailored for HAR systems that recognize concurrent activities. Their approach involves using boosting algorithms to rank the various tags associated with a sample.

**Table 4. Different types of complex activities in HAR alongside their solutions.**

| Type | Example | Solutions |
|---|---|---|
| Composite | Cooking, Working, Assembling | LSTM networks [82] |
| Concurrent | Talking and eating at the same time | Multi-label classification [83]<br>LSTM multilayer networks [84] |
| Multi-occupant | Performs by a group of users like playing football | Using motion and environmental sensors in parallel [86] |

The composite activities include simple activities set in sequence form, resulting in high-level activities such as cooking, working, and assembling. Therefore, detecting such activities requires environmental information and is more complex than simple activity recognition. The authors also [82] proposed the CNN models

This ranking is achieved by integrating sub-optimal learning models guided by a voting system. Game theory was then applied to calculate the final loss function of their proposed method, optimizing the outcome. Similar studies [88] combined several convolutional recurrent neural networks (CNN) to identify music genres and related labels. The compared individual CNN networks were based on similarity criteria to obtain the best results (genres close to the intended music) and using Graph Transduction Game (GTG) method. The GTG is a non-cooperative game in a class of semi-supervised learning techniques to estimate the classification function over the labeled and unlabelled data [89] that can be reciprocally utilized to solve the multi-label problem in complex activities.

To apply GTG to the problem of complex activities, let $X = \{x_1, x_2, \ldots, x_n\}$ be a set of $n$ data points, where each data point represents a frame of video or sensor data. Let $Y = \{y_1, y_2, \ldots, y_l\}$ be a set of l activity labels, each representing a different activity that could be performed in a given frame. The goal is to predict the labels of each data point, which is a multi-label classification problem.

To solve this problem using GTG, we first construct a graph $G=(V,E)$, where $V$ is the set of nodes representing the data points and $Y$ is the set of labels, and $E$ is the set of edges representing the relationships between the data points and the labels. Specifically, we use a weighted graph where the edge weights represent data points' similarity and the labels' relationships.

of the labels, subject to the constraints imposed by the graph structure and the available data.

We use a game-theoretic approach to achieve this by defining two players: the labeler and the transducer. The labeler selects a labeling function f based on the available data and the current state of the graph, while the transducer adjusts the labeling function based on the relationships between the labels and the data points. The game continues until a stable labeling function $f^*$ is reached.

To formalize the game, let $L(f)$ be the loss function that measures the error between the predicted and actual labels, and let $R(f)$ be the regularization function that imposes constraints on the labeling function f. The labeler tries to minimize $L(f)+R(f)$, while the transducer tries to maximize $L(f)-R(f)$. We can write this as a two-player zero-sum game:

$$\min_f \max_g L(f) - R(f) + G(g,f) \quad (14)$$

where g is the transducer's strategy, $G$ is the game matrix that encodes the relationship between the labeler's and transducer's strategies, and the min-max operator represents the equilibrium solution of the game.

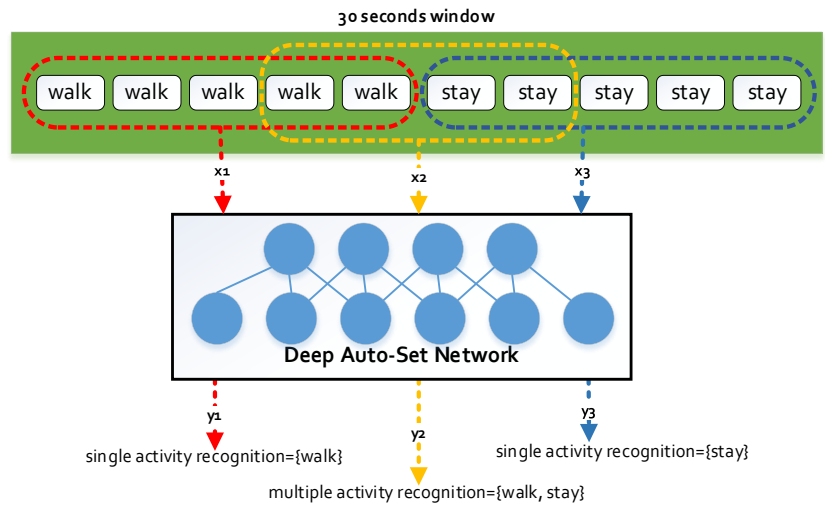

**Figure 9. An overview of multi-label classification architecture for HAR systems with a long-time window.**

We then define a labeling function $f : V \times Y \to \{0,1\}$ that assigns a binary value to each pair of data points and label, indicating whether or not the label applies to the data point. The goal is to find a labeling function $f^*$ that maximizes the accuracy

**4.8. Data Segmentation**

Due to the input of motion sensors data streams in HAR systems, a fixed length usually is applied to segment time windows, which is considered a limitation. One or several activities in a one-time window is possible regarding the fixed time windows in the best case. A label is referred to the entire time window to tag this kind of data that might not be true for all cases. The minor windows cannot extract much information; however, longer-time windows extract more comprehensive information, resulting in more complexity (due to multiple activities). As shown in Figure 9, A hierarchical view is proposed to select the optimal time windows that start from a long time window, and these time windows become so narrow that only one activity appears in each window [90].

In the hierarchical approach, fine-grained classifications are applied in the first "two-class classification" level if two consecutive windows have different labels that become less than the threshold, besides the second "two-class classification," level accrues by primary level decisions.

choosing a particular label based on the payoffs of all players in the game. To apply the MPGL method to sensor-based HAR, the labeled data is used to train a classifier, which is then used to label the unlabelled data. The MPGL game is then played on the labeled and newly labeled data to improve the accuracy of the labels.

As observed in Urbani's thesis, their proposed method gets satisfactory results in recognizing related music genres, and we believe that their idea can be helpful on HAR due to the nature of time-series data for both systems.

The possible solution can be defined as Let $X = \{x_1, x_2, ..., x_n\}$ be the set of all data segments, and $Y = \{y_1, y_2, ..., y_k\}$ be the set of all possible labels.

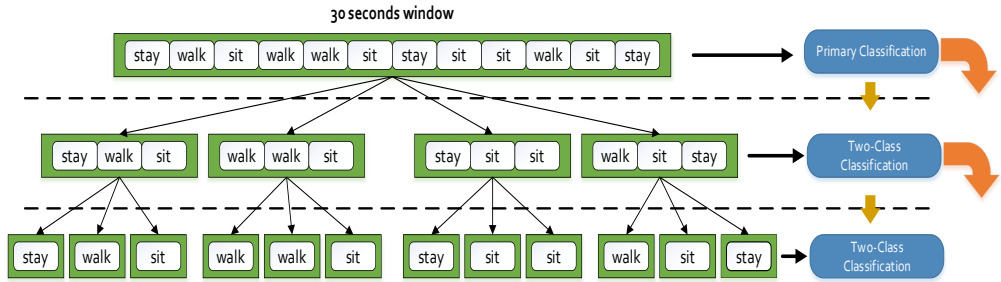

**Figure 10.** An illustration of a hierarchical view to extract more information from a long-time window in HAR systems.

Two-class classifiers only search for the activities that are appeared in the primary classification. According to Figure 10, the data segmentation challenge can also be defined as a multi-label problem. Urbani's idea can also be implemented as a game theory-based data segmentation technique for HAR. We suggest a supervised method based on the non-cooperative multi-player game computed by a dynamic replicator that can be used for sensor-based HAR called the multi-player game-theoretic framework for labeling sensor data (MPGL). This method can spread labeled data to unlabelled samples and find the most similar labels to segmented data.

The MPGL method is based on the concept of a non-cooperative multi-player game, where each player represents a sensor and the game is played to decide the label of a particular segment of data. The game's goal is for each player to maximize their payoff, which is a function of the accuracy of their labeling decision.

The game is played iteratively, with each player making a decision based on the decisions of their neighbors in the sensor network. The decisions are updated using a dynamic replicator equation, which updates the probabilities of each player

Let $L = \{l_1, l_2, ..., l_m\}$ be the set of all labeled data segments, where each segment $x_i$ in $L$ is associated with a label $y_j$ in $Y$. We define a player $i$ as a sensor in the network and let $P_i(y_j)$ be the probability that player $i$ assigns to label $y_j$. Let $H_i(x_i, y_j)$ be the accuracy of player $i$ when labeling data segment $x_i$ with label $y_i$.

The goal of the game is for each player $i$ to choose a label $y_j$ that maximizes their payoff. The payoff of player $i$ when choosing a label $y_j$ is defined as follows:

$$Payoff_i(y_j) = \frac{1}{n} sum_{x_k\ in\ X}\ H_i(x_k, y_j) P_j(y_j) \quad (15)$$

where the sum is over all data segments $x_k$ in $X$, and $P_j(y_j)$ is the probability that player $j$ assigns to label $y_j$.

The game is played iteratively, with each player $i$ choosing a label $y_j$ that maximizes their payoff. The probabilities of each player choosing a particular label are updated using a dynamic replicator equation:

$$\frac{dP_i(y_j)}{dt} = P_i(y_j)(Payoff_i(y_j) - lambda_i) \quad (16)$$

where $lambda_i$ is a constant that controls the overall intensity of the game.

To apply the MPGL method to sensor-based HAR, we first use the labeled data $L$ to train a classifier that maps data segments to labels. We then use the classifier to label the unlabelled data segments. The MPGL game is then played on the labeled and newly labeled data to improve the accuracy of the labels.

**4.9. Computation costs**

The computation costs are also another fundamental challenge for HAR. The learning models are needed to be lightweight calculations in a device with low resources; since the memory and processing resources are costly and require a high-power consumption. In the simplest case, taking advantage of shallow learners with efficient hand-crafted features is recommended, but various solutions are proposed in the literature. However, extracting robust hand-crafted features is not possible in many cases. Therefore, deep learning was proposed alongside combined hand-crafted features alongside deep models [91].

The quantized learning models are also another efficient solution. In these models, all network weights and input data are limited to 1 to -1, and simplifying input samples from continuous to discrete reduces input calculations [92]. In addition to lightweight models and model quantization techniques, knowledge distillation (KD) is proposed to reduce the computation costs in learning models. In KD, a small and lightweight model mimics its behavior from a pre-trained model learned by a large dataset. The presented learning model reduces training computation costs due to prior knowledge [93].

In this regard, Wang *et al*. [94] developed a GAN-based KD method based on the game theory to reduce the computation cost of the HAR systems. They designed a three-player game called KDGAN with the components of a classifier, a trainer, and a discriminator; the trainer and classifier train each other due to distillation, and adversarial losses also train the adversarial model. They also optimized distillation and adversarial lead to the classifier model's training, distributing the correct data in equilibrium.

As a possible solution to solve the computation costs problem for HAR, cooperative games involve training multiple models to work cooperatively to solve a problem. Each model is responsible for recognizing a subset of the activities, and the models communicate with each other to share their knowledge and make better predictions.

We can develop a cooperative game with the KD framework to combine these two techniques for HAR. In this framework, we first train a large and complex model on the entire dataset, which serves as the teacher model. We then train multiple smaller student models, each responsible for recognizing a subset of the activities. These student models communicate with each other to share their knowledge, and they are trained using KD to transfer the knowledge from the teacher model to the student models.

During training, we can use a cooperative game approach where each student model is rewarded for correctly recognizing its subset of activities, and the models cooperate to maximize the system's overall accuracy. This approach can lead to better performance than training a single model on the entire dataset while being more efficient and lightweight than the large teacher model.

**4.10. Privacy**

Monitoring users' daily activities is HAR systems' main procedure, which conflicts with users' privacy. Since the collected data from different motion sensors can extract information such as age, gender, height, and weight, various practical solutions are proposed in the literature to trade-off between utility and privacy. For instance, they were embedding a multi or single-objective adversarial loss function in learning the procedure for user trait recognition in which all loss values of each user trait tend to minimize [95][96][97]. However, they are limited to protecting only pre-defined user traits in training steps. The user privacy protectors usually act as agents on local devices to anonymize sensitive information before releasing it in a shared space (e.g., cloud space). They covert the sensitive users' information to a random and noisy style except for the foreground data [98].

In game-theory-based methods, Miyaji and Rahman [99] developed a privacy model based on cryptographic hypotheses and provided a secure and efficient set-intersection protocol with game theory.

Their findings indicate that the proposed protocol satisfies the tight Nash equilibrium computational versions and is compatible with trembles. On the other hand, Huang [100] proposed an efficient three-party game framework for mutual interaction between users, platforms, and adversaries. In similar studies, model designers usually add noisy data to users' sensitive information for users' information protection, but their context-aware quality service was lost.

Therefore, the game theory can balance user privacy and quality service as a powerful tool. Later, Huang extended his framework to a multi-user scenario with a two-layer and three-party game framework that is more realistic by exploring the interaction between users and different parties. Inspired by [100], we believe that Huang's idea could be an exciting research opportunity, and a possible formulation for a new data anonymization mechanism based on game theory for HAR to address privacy challenges is:

Let the HAR dataset be denoted by $D$, where $D = \{(x_1, y_1), (x_2, y_2), ..., (x_n, y_n)\}$ represents $n$ instances of sensor data $(x_i)$ and corresponding labels $(y_i)$.

Let $S = \{s_1, s_2, ..., s_m\}$ be a set of $m$ anonymized datasets, where each $s_j$ is a permutation of $D$ such that the labels are shuffled randomly. Let $P$ be the set of all possible player types, where each player type $p_i$ is associated with a probability distribution over the set S, denoted by $P_i = \{p_{\{i,1\}}, p_{\{i,2\}}, ..., p_{\{i,m\}}\}$. Each player i selects a strategy $s_i$ from the set $S$, based on its player type and the strategies selected by other players, the expected utility of player $i$ is maximized.

The utility function for each player $i$ is defined as $U_i(s_1, s_2, ..., s_k) = ACC_i(s_i)$ the accuracy of the player i's HAR model trained on the dataset $s_i$. The accuracy is computed as the fraction of correctly predicted labels on a held-out test set. The game is played in rounds, where in each round, each player $i$ selects a strategy $s_i$ based on its player type and the strategies selected by other players and then computes its expected utility $U_i(s_1, s_2, ..., s_k)$ based on the current strategies.

The game converges to a Nash equilibrium, where no player can improve its utility by unilaterally changing its strategy.

The anonymized dataset selected by each player at the Nash equilibrium is used for training the HAR models, and the corresponding labels are released for public use. The game-theoretic mechanism ensures that each player's privacy is protected, as no player can infer the true labels of the instances in the dataset. At the same time, the mechanism incentivizes players to select high-utility anonymized datasets, leading to better accuracy of the trained HAR models.

### 4.11. Interpretability

Most data may appear in an unrelated time window or carry noisy data during a specific activity's sampling phase. For instance, in the diagnosis of Parkinson's disease, the occurred event appears at a specific time (when the user is standing), and the system only pays attention to that part of the data. The interpretable deep learning models are presented to overcome mentioned challenges, automatically deciding whether to pay more attention to the input data [101].

The attention mechanism distinguishes certain parts of extensive sample data during interpretation. The suggested mechanism is divided into soft and complex forms according to the differentiation role in the input data. In soft attention, the values 0 to 1 refer to the input of each system input element, and then the system decides which elements to be the center of attention [102] [103] [101].

In contrast to soft attention, different sections are interpreted simultaneously by zoning a specific sample instead of examining the input elements in the data interpretability for hard attention. The model gradually decides to pay more attention to one part of the system, and the knowledge gained in the following steps is intertwined into other regions for interpretability. Deep Q-Learning and reinforcement learning were employed in the decision-making procedure of attention mechanisms to select essential parts of data. Also, some other studies were conducted based on multi-agent systems in the HAR field [104] [105] [106].

In addition to identifying essential locations in an input data sample, the interpretability mechanism can differentiate and mark the micro-activities in an input sample [107].

Many studies have applied attention mechanisms to optimize learning problems based on game theory. In this regard, Ethayarajh and Jurafsky [108] found a logical relationship between shapely values and attention flows post-processing procedures and proved that attention weights and leave-one-out values (an erasure-based method) could not be placed in shapely values.

They used attention-flow techniques to analyze the input data of a learning system to interpret the shapely values as a collaborative problem.

However, their method is applied only to a specific function of the issues that did not guarantee the accuracy of a predictive model. In another study, Tripodi and Navigli [108] developed an integrated framework to solve word sense disambiguation (WSD) problems with game theory and the attention mechanism. They used vague words as players and the sense of the words as each player's strategy in a non-cooperative game, which resulted in an attention mechanism on the input samples of the WSD learning problem. The results indicated acceptable performance compared to the state-of-the-art studies in this research field that their proposed method may help to attention mechanism problems in HAR.

This section presented a literature review of eleven main HAR challenges in separated sub-sections. Then, we discussed the game-theory-based methods for each challenge and suggested possible research opportunities based on the game theory. In the next section, we also summarize future research directions that can be helpful to HAR systems based on game theory.

## 5. Future Research Directions

The HAR is investigated using various dimensions and approaches, and many practical articles have been published on operational solutions to address the main challenges in recent decades. However, the HAR models have not been evaluated from a game theory perspective; therefore, the present study attempted to identify similar game-theory-based machine learning methods to address the elected challenges of our paper. The study aimed to identify new research opportunities in the HAR field, considering solving the problems based on game theory solutions. HAR provides new research directions and opportunities based on non-cooperative games.

### 5.1. Noise filtering with non-cooperative games

As the filtering methods improve the HAR systems' quality and game theory tools could increase the efficiency of noise filtering tasks, the proposed method for noise reduction in sensor-based HAR involves modeling the interaction between sensor data and noise as a non-cooperative game between two players: the sensor data player and the noise player.

The sensor data player aims to classify human activity correctly, while the noise player aims to introduce noise to mislead the sensor data player. A Nash equilibrium is computed to find the optimal solution, representing a set of strategies where neither player can improve their payoff by unilaterally changing their strategy. The sensor data player uses this stable solution to classify human activity while mitigating the noise introduced by the noise player, leading to improved classification accuracy even in noisy environments. However, the approach has some limitations, including the assumption that the noise player's strategy is known and the computational expense of finding the Nash equilibrium, especially for large datasets and complex models. Further research is necessary to explore the effectiveness and scalability of this approach for real-world HAR applications.

### 5.2. Using game theory to identify reliable sensors in feature extraction

In the paper [58], efficient sensors from the input data of a learning model were identified using non-cooperative games and automata machine learning. They divided the input sensors into reliable and unreliable categories and investigated sensor efficiency with automata machine learning. The findings indicated that the proposed method could be applied in HAR systems due to an extensive range of sensors that obtain the best fusion. In the proposed solution, using multiple sensor modalities has several advantages, including improved accuracy by capturing more comprehensive and robust features, increased system robustness by reducing the impact of sensor failures or errors, flexibility in adapting to different contexts and use cases, and generalizability by learning more generalizable features.

However, there are also limitations such as increased complexity in processing and storage of high-dimensional data, data synchronization challenges when dealing with sensors with different sampling rates or latencies, privacy concerns due to the potential compromise of sensitive information, and limited interpretability that can limit users' ability to understand and trust the system.

### 5.3. Improving efficiency of data annotation

According to the proposed method by [70], the human error rate makes the data annotation procedure less accurate, or part of the data may remain without labels. The game theory-based adversarial systems produce artificial samples to make the system compatible and resistant to worst-case perturbation.

Reducing the error of annotation of data is probable using the proposed logic. As described in section 4.3, the annotation procedure is a fundamental challenge that machine-learning-based systems face, especially HAR. Researchers recently developed different data annotation mechanisms without human intervention or semi-autonomous operations.

Also combining the game theory tools presented in the literature can provide excellent results in reducing the data annotation error rates.

The proposed CROWDGAME-based data labeling system offers several advantages for HAR. First, it enables the augmentation of a dataset with additional accurately labeled data tuples, even when labeled data is scarce. Secondly, it leverages multiple players' collective knowledge and expertise to improve the accuracy and reliability of labeled data. Thirdly, it is a flexible and scalable approach that can accommodate a large number of players and can be easily integrated into existing labeling pipelines.

However, the proposed method also has some limitations. One potential issue is the potential for bias in the labeling process, as the accuracy of the labeled data depends on the quality of the rules generated by the players. Additionally, the system may be vulnerable to strategic behavior by players, as they may be incentivized to generate rules that favor their interests rather than accurately labeling the data. Finally, the system relies on the availability of many players, which may not always be feasible in practice. Nonetheless, with proper safeguards and careful design, the proposed CROWDGAME-based data labeling system can be an effective and efficient approach for improving the accuracy and reliability of labeled data in HAR.

### 5.4. Solving classes imbalance problem with game theory

In everyday life, routine activities are widespread, and other activities occasionally make unbalanced classes in HAR. Accordingly, some activities may contain fewer samples, reducing the HAR model performances. Many related studies were conducted to overcome class imbalance problems, such as generating artificial data. However, a literature review shows that methods based on game theory have also provided well-improvement for class imbalance data.

For example, researchers in [13] developed a non-collaborative game based on a filtering system for synthetic labeling data, and combining previous HAR methods for balancing imbalance classes with game theory can be exciting for new research directions.

However, the proposed method may also have limitations, such as the potential for overfitting, increased complexity, and the assumptions of game theory. Therefore, these limitations should be carefully considered before implementing this method.

### 5.5. Solving distribution discrepancy problems using game-theory-based transfer learning approaches

Transfer learning is the practical technique to solve distribution discrepancies in the user and time directions of HAR systems. As we studied the distribution discrepancy problem for HAR in section 4.6, they appeared in the three principles: distribution discrepancy in the user axis, time, and sensor.

Then we observed that a general transfer learning method based on game theory had been proposed [81] that can be useful for HAR. The authors [81] used a minimax game to develop an optimal transfer learning system that finds the best data sources for training the destination model.

In our suggestions, a new method for selecting the best source data in the axial distribution discrepancy challenge assisted in transfer learning for HAR with a mini-max game.

The method for selecting the best source data in the axial distribution discrepancy challenge has several advantages that first allow for the selection of high-quality source data, which is most helpful in improving the performance of the HAR model. Additionally, the mini-max game component of the approach helps to ensure that the selected data is representative of the overall distribution of data and not biased towards any particular subset. However, the method also has some limitations. One potential drawback is that it requires a significant amount of computational resources, which may make it less accessible to researchers or practitioners without access to high-performance computing.

Additionally, the method may not always be effective in cases where the source data is highly variable or noisy, making it challenging to identify the most valuable subsets for transfer learning.

### 6.6. Recognizing concurrent activities using game theory

Concurrent activity recognition can be defined as a multi-label data problem. According to Section 4.7, extensive research was applied to using binary classifiers to recognize the simple activities associative in a concurrent data sample. However, they suffered from high computation costs. Therefore, some studies applied the multi-layered LSTM (each layer for a particular action), indicating high accuracy but not a flexible system. In [88], a synthetic learning method based on CRNN networks was presented using GTG games to extract music genres that identify and track concurrent activities. Given that both HAR and genre identification research fields are as follows as using time-series data, we believe that [88] can perform a satisfactory improvement in the problem ahead.

Advantages of using GTG for sensor-based HAR for complex activities include: 1) GTG is a semi-supervised learning method, meaning it can learn from labeled and unlabeled data. It is helpful in HAR because labeled data can be expensive and time-consuming, and many unlabeled examples may be available. 2) GTG can handle complex activities involving multiple sensor inputs and may have high variability. It is because the graph structure can capture the relationships between different sensor inputs and help identify patterns in the data. 3) GTG is robust to noisy data because it can use the graph structure to smooth out noisy sensor data and handle missing data.

However, there are also limitations to using GTG for sensor-based HAR for complex activities, including

1) GTG requires domain expertise to design the graph structure and to interpret the results. It may be a limitation for non-experts who cannot effectively design the graph or analyze the results.
2) GTG may not scale well to large datasets or complex graphs. The method's computational complexity increases as the graph's size and the number of labeled nodes increases.
3) The performance of GTG is sensitive to the choice of graph structure and the parameters used to construct the graph. It may require careful tuning to achieve optimal results.

**5.7. Using game-theory-based knowledge distillation to improve computation cost**

The HAR systems are usually implemented on devices with low processing power; hence, optimizing their computation cost is mandatory. In previous works, shallow learners using the hand-crafted feature, model quantization, transfer learning or KD was applied to reduce the computation cost. The researchers have never used game theory to improve the performance of these methods in HAR. At the same time, the paper in [94] shows that applying game theory can help increase performance and reduce computing costs. They presented a game-theory-based KD to save HAR learning systems' computation cost.

In our suggestion, combining KD and cooperative games could be a practical approach to developing efficient and lightweight models for HAR. By transferring the knowledge from a larger model to smaller models that work cooperatively, we can achieve high accuracy using less computational resources and memory. However, there are still some limitations that should be considered.

First, the performance of the student models may not be as good as that of the teacher model, especially if the teacher model is extensive and complex. That is because KD involves transferring knowledge from a larger and more complex model to a smaller and simpler model, which may result in some loss of accuracy. Therefore, it is essential to carefully balance the size and complexity of the teacher and student models to achieve the desired trade-off between accuracy and efficiency.

Second, the cooperative game approach may require significant coordination and communication among the student models, which can add computational overhead and complexity. The models must be trained to work together effectively and efficiently, and the communication protocol must be designed carefully to ensure that the models share relevant and valuable information.

Finally, the choice of sensor data and feature representation can also affect the performance of the models. Different sensors and features may be more or less suitable for different types of activities. It may be necessary to experiment with different combinations to find the best approach for a particular application.

**5.8. Proposing a new data anonymization mechanism to protect users' sensitive information with game theory**

In reference [99], a regulatory protocol, grounded in game theory, was devised to safeguard users' sensitive information, presenting both a safe and effective solution. Similarly, research outlined in [100] suggested a game-theory-centric strategy that serves to protect user privacy. This approach functions as a context-aware service, striking a balance between maintaining user privacy and the provision of quality services. As indicated by these studies, game theory has demonstrated success in achieving equilibrium in user privacy utility, a development which holds exciting research prospects for HAR.

Several potential advantages are associated with this approach, including: the provision of a robust privacy guarantee, enhanced robustness, scalability, and the flexibility of the HAR model. However, it is essential to also consider potential limitations.

Table 5. Proposed non-cooperative game-based analytical tools for HAR challenges alongside the limitations.

| HAR Challenge | Game Tool | Idea | Limitations | References |
|---|---|---|---|---|
| Noise filtering | Strategic Form | Filtering the noisy data of motion sensors using a non-cooperative game as a decision-maker | Dependency of prior information to control thresholds by user; Distribution discrepancy problem in input resources; Low performance in real-time systems | [10] [117] |
| Feature extraction | Strategic Form | Finding reliable sensors with non-cooperative games and automata machine learning | High computation cost; Low performance in real-time systems; Sensitive to class imbalance problem | [58] |
| Annotation scarcity | Minimax Strategy | Improving the efficiency of data annotation operations with crowd games, adversarial systems, and minimax strategy | Do not take advantage of unlabelled data; Low performance in imbalanced data | [70] |
| Class imbalance | Strategic Form | Solving imbalanced classes by using a non-cooperative game | Over-generalization problem | [13] |
| Distribution discrepancy | Minimax Strategy | Solving distribution discrepancy problem using transfer learning and Mini-Max game | Retraining process for each new user | [81] |
| Concurrent activities | Graph Transduction Game | Recognizing concurrent activities based on the GTG method | Low performance in multi-class classification; High computation cost | [88] |
| Computation cost | Minimax Strategy | Using game-theory-based knowledge distillation to optimize computation cost | Dependent to hyperparameters; Low performance in high-variance gradient updates | [94] |
| User privacy | Non-cooperative | Balancing user privacy and utility with game theory approaches | Low performance in real-time systems; Mismatching between attention and continuous attention | [99] [100] |

1) Game-theoretic approaches may not universally apply to all data types or scenarios. For instance, these methods might prove ineffective if the data is overly noisy or if the attacker possesses extensive prior knowledge about the data.
Game-theoretic approaches inherently involve complexity, potentially requiring significant expertise for effective implementation and usage, which could present challenges for those lacking specialized knowledge.
2) There might exist trade-offs between privacy and utility when employing game-theoretic methodologies. Specifically, enhancing privacy might lead to a reduction in data accuracy or usability.
3) Game-theoretic strategies often hinge upon presumptions about attacker behavior, which may not consistently align with real-world scenarios. This could potentially curtail the efficacy of these approaches in certain situations. Upon reviewing literature on HAR systems and game theory solutions, new research directions have emerged, pointing towards addressing challenges in HAR through the application of game theory analytical tools.
Subsequently, we will present a summary of all identified research opportunities in Table 5.

## 6. Conclusion

This review paper explored the potential of game theory solutions in sensor-based human activity recognition (HAR). It highlighted their benefits in improving performance, model efficiency, and recognition algorithms and addressing common challenges in the field such as data scarcity, model interpretability, and real-time performance. We presented different approaches to using game theory in sensor-based HAR including non-cooperative and cooperative games, and discussed the advantages and disadvantages of each approach. We also proposed new approaches for using game theory in sensor-based HAR and evaluated their effectiveness compared to traditional approaches. Through our analysis of the current literature, we found that game theory solutions have the potential to enhance the accuracy and robustness of HAR models significantly.

Future research should focus on developing more effective and efficient recognition systems using game theory concepts.

Our review has contributed to bridging the gap between game theory and HAR research, providing valuable insights for the researchers in the field and guiding future research efforts.

We hope our work will inspire further exploration of game theory solutions in sensor-based HAR, ultimately leading to more accurate and efficient recognition systems that can benefit various domains, including healthcare, sports, and security.

# مروری بر راه‌حل‌های نظریه بازی‌ها در تشخیص فعالیت‌های انسانی مبتنی بر حسگر

محمدحسین شایسته، بهروز شاهرخ زاده* و بهروز معصومی

گروه مهندسی کامپیوتر و فناوری اطلاعات، واحد قزوین، دانشگاه آزاد اسلامی، قزوین ، ایران.

**چکیده:**

سیستم‌های تشخیص فعالیت انسان به طور خودکار فعالیت‌های انسانی را با استفاده از داده‌های حسگر ها شناسایی می‌کنند که کاربردهای متعددی در مراقبت‌های بهداشتی، ورزش، امنیت و سایر تعاملات انسان با رایانه دارد. با وجود پیشرفت‌های قابل توجه در تشخیص فعالیت انسان، همچنان چالش‌های بسیاری در این حوزه وجود دارد. تئوری بازی‌ها به عنوان راه‌حلی مناسب برای مواجهه با این چالش‌ها در مسائل یادگیری ماشین به خدمت گرفته شده است. اما، تاکنون کار تحقیقاتی جامعی درباره استفاده از راه حل های تئوری بازی‌ها برای مسائل تشخیص فعالیت انسان ارائه نشده است. این مقاله بررسی می‌کند که چگونه تئوری بازی‌ها به عنوان راه‌حل هایی کارآمد برای حل مشکلات سیستم‌های تشخیص فعالیت انسان استفاده می‌شود و با پیشنهاد رویکردهای جدید نظریه بازی برای حل مسائل تشخیص فعالیت انسان، شکاف بین نظریه بازی و این حوزه تحقیقاتی را پر می‌کند. مشارکت‌های اصلی این کار شامل بررسی چگونگی بهبود دقت و استحکام مدل‌های تشخیص فعالیت انسان با استفاده از تئوری بازی، بهینه سازی الگوریتم‌های تشخیص با استفاده از مفاهیم تئوری بازی و بحث در مورد رویکردهای تئوری بازی در مقایسه با روش‌های موجود تشخیص فعالیت انسان است. هدف این مقاله ارائه پیش‌بینی در مورد پتانسیل تئوری بازی به عنوان راه‌حلی برای تشخیص فعالیت انسان بر اساس حسگرها و کمک به توسعه یک سیستم تشخیص دقیق‌تر و کارآمدتر در تحقیقات آینده است.

**کلمات کلیدی:** تشخیص فعالیت انسان، تئوری بازی‌ها، یادگیری ماشین، یادگیری عمیق، چالش‌ها، راه حل ها، فرصت‌های تحقیقاتی.